\documentclass[aps,superscriptaddress,prb,amsmath,amssymb,reprint]{revtex4-2}

\usepackage{hyperref}
\usepackage{graphicx}
\usepackage{mhchem}
\usepackage{amsmath}
\usepackage{xcolor}
\usepackage{array}
\usepackage[english]{babel}
\usepackage{tabularx}
\usepackage{multirow}
\usepackage{soul}
\usepackage{rumathgrk} 
\usepackage{verbatim}
\usepackage{lipsum}

\newcommand{\degree}{^{\circ}}
\newcommand{\orcid}[1]{\href{https://orcid.org/#1}{\includegraphics[scale=0.45]{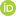}}}

\begin{document} 
	
	\title{Kerr effect induced by exchange interaction of electrons separated by a tunnel barrier in a double quantum well}
	
	\author{V.~K.~Kalevich~\orcid{0000-0002-8010-2714}}
	\email{Corresponding author: kalevich@solid.ioffe.ru}
	\affiliation{Ioffe Institute, Russian Academy of Sciences, St.\,Petersburg 194021, Russia}
	
	\author{K.~V.~Kavokin~\orcid{0000-0002-0047-5706}}
	\affiliation{St.\,Petersburg State University, St.\,Petersburg 198504, Russia}
	
	\author{M.~M.~Afanasiev~\orcid{0000-0003-3955-8389}}
	\affiliation{Ioffe Institute, Russian Academy of Sciences, St.\,Petersburg 194021, Russia}
	
	\author{B.~F.~Gribakin~\orcid{0000-0002-5339-1813}}
	\affiliation{St.\,Petersburg State University, St.\,Petersburg 198504, Russia}
	\affiliation{Laboratoire Charles Coulomb, UMR 5221 CNRS-Universit\'e de Montpellier, F-34095 Montpellier, France}
	
	\author{M.~I.~Kuzmenko~\orcid{0009-0003-4004-4737}}
	\affiliation{Ioffe Institute, Russian Academy of Sciences, St.\,Petersburg 194021, Russia}
	
	\author{G.~Karczewski~\orcid{0000-0003-3441-1425}}
    \affiliation{Institute of Physics, Polish Academy of Sciences, Warsaw 02668, Poland}
	
	\author{Yu.~G.~Kusrayev~\orcid{0000-0002-3988-6406}}
	\affiliation{Ioffe Institute, Russian Academy of Sciences, St.\,Petersburg 194021, Russia}
	
	\begin{abstract}
		In a structure with two tunnel\nobreakdash-coupled quantum wells of different widths, the spin dynamics resulting from resonant pulsed optical pumping of the narrow-well exciton includes the wide-well electron magnetization dynamics. Our analysis shows that the effect is driven by electron exchange between narrow-well excitons and spin-polarized electrons in the wide well. A theoretical model of the spin Kerr effect has been developed accounting for the interwell electron spin exchange. In the studied double-well structure with CdTe and Cd$_{0.98}$Mn$_{0.02}$Te quantum wells and a well\nobreakdash-separating barrier thickness of 5 monolayers (1.6\,nm), the model accurately describes the experimental results and allows us to estimate the interwell electron exchange constant as $\delta_{e} \approx 0.9\times10^{-15}~\textrm{eV~cm}^{2}$.
	\end{abstract}

	\maketitle
	
	\section{Introduction}
	
	Exchange interaction of electrons, having a significant effect on the physical properties of semiconductors and semiconductor structures, leads to a number of striking physical phenomena and is important for modern technologies \cite{Anderson1963, Kondo1968, Konstantinov1997}. For example, the exchange interaction of bright excitons with a reservoir of dark ones with the wave vectors outside of the light cone in a quantum well (QW) under optical pumping in a magnetic field in the Voigt geometry induces periodic oscillations in the intensity of bright exciton luminescence\,\cite{Trifon2019}. Numerous nontrivial effects are demonstrated by hybrid structures containing layers of magnetic diluted semiconductors, including structures with tunnel-coupled QWs~\cite{DMS, Scalbert}.
	
	Our recent studies of spin dynamics in tunnel-coupled quantum wells CdTe and Cd$_{0.98}$Mn$_{0.02}$Te, which differ in width, have revealed that the spin dynamics of excitons in the narrow well, occurring under their resonant pulse optical pumping, includes the dynamics of electron magnetization in the wide well, although the electron level in the wide well is significantly, by 55\,meV, lower than that in the narrow one~\cite{PSS2023}. The measurements were performed using the spin Kerr effect in the pump-probe mode. Analysis of the obtained results allowed us to assume that this effect is caused by the exchange interaction of spin-polarized electrons in the wide well with that bound into excitons in the narrow one~\cite{PSS2023}.
	
	This paper presents the spin Kerr effect model, which we developed for describing the discovered effect. Comparing it with experimental data, it allows us to estimate the electron interwell exchange constant. For the tunnel barrier with a thickness of 1.6\,nm (5~monolayers, MLs) separating CdTe and Cd$_{0.98}$Mn$_{0.02}$Te wells, its value is $\approx0.9\times10^{-15}~\textrm{eV\,cm}^{2}$.
	
	The paper is organized as follows. Section~II gives an overview of the studied samples, the experimental setup, and data from our recent experiments of observing the discovered spin Kerr effect, which, after processing, are presented in the form convenient to compare with theory. In Sec.~III we discuss a calculation of the wave functions and energy levels in the investigated double quantum well and a theoretical estimate of the electron-electron exchange interaction constant. Section~IV presents our proposed model of the Kerr effect in the narrow QW in a magnetic field in Voigt geometry, which includes spin exchange between electrons in neighboring wells, as well as reflection of a light wave from a heterointerface with a substrate. This model is used to approximate the experimental dependences of the Kerr effect and to find the value of the electron-electron exchange constant from the experiment. A brief summary of the obtained results is given in Sec.~V. In the Appendices the expression for the Kerr effect is obtained, including electron\nobreakdash-hole exchange interaction in exciton, and it is shown that this interaction can be ignored in the studied semimagnetic QW (Appendix~A); with use of the expression for the reflectance of the entire structure, which takes into account reflection from the substrate, an approximation of the experimentally measured reflection spectrum is carried out and the reflection parameters are found (Appendix~B); using the magnetooptical Kerr effect in an alternating magnetic field in Faraday geometry, the parameters of exciton resonance in the narrow well and the tunneling time of charge carriers from the narrow well were measured (Appendix~C).
	
	\section{Experimental}
	
	\subsection{Studied structure.}
	
\begin{figure}[t]
		\includegraphics[width=0.9\linewidth]{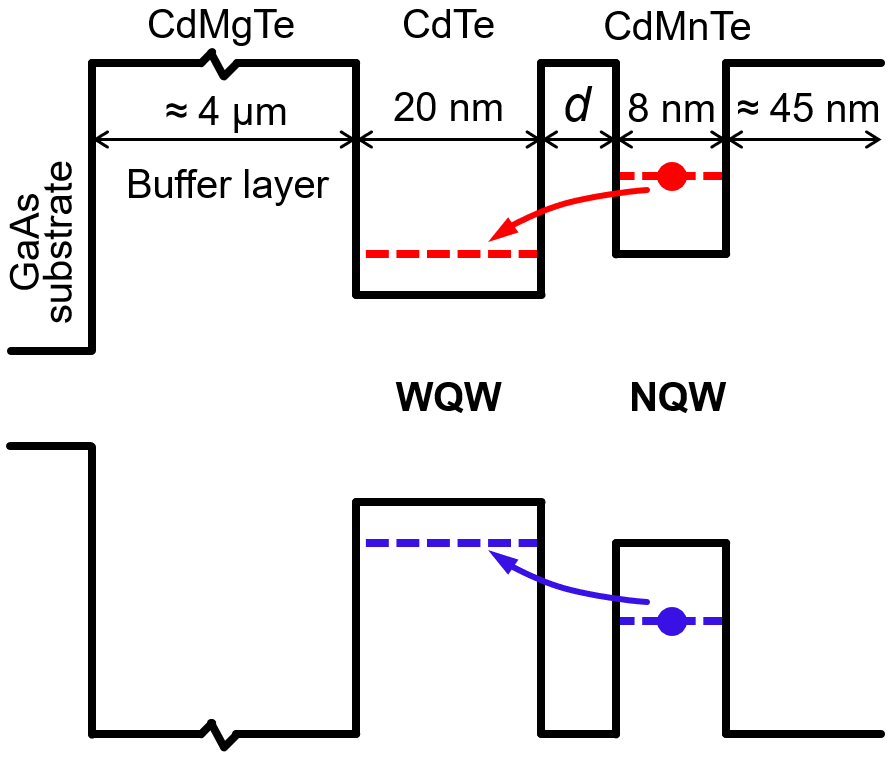}
		\caption{Schematic energy diagram of the studied DQW.}
		\label{fig:structure}
	\end{figure}

	The double quantum well (DQW) under study consists of wide (20\,nm) nonmagnetic CdTe quantum well (WQW) and narrow (8\,nm) magnetic Cd$_{0.98}$Mn$_{0.02}$Te quantum well (NQW) separated by a nonmagnetic Cd$_{0.88}$Mg$_{0.12}$Te spacer of thickness $d=5$~MLs (1.6\,nm), see Fig.\,\ref{fig:structure}. It was grown by molecular-beam epitaxy on a (100)-GaAs substrate. To reduce the strain induced by II-VI on III-V heteroepitaxy the GaAs substrate was overgrown by a Cd$_{0.88}$Mg$_{0.12}$Te buffer layer of $\approx$\,4\,$\muup$m width. The NQW, and thus the entire structure, was capped by a Cd$_{0.88}$Mg$_{0.12}$Te layer of $\approx$\,45\,nm width. All grown layers were not intentionally doped. During the growth process, the wafer was not rotated, so the cross-sections of the layers have a weakly expressed wedge shape.
	
	\begin{figure}[b]
		\includegraphics[width=0.95\linewidth]{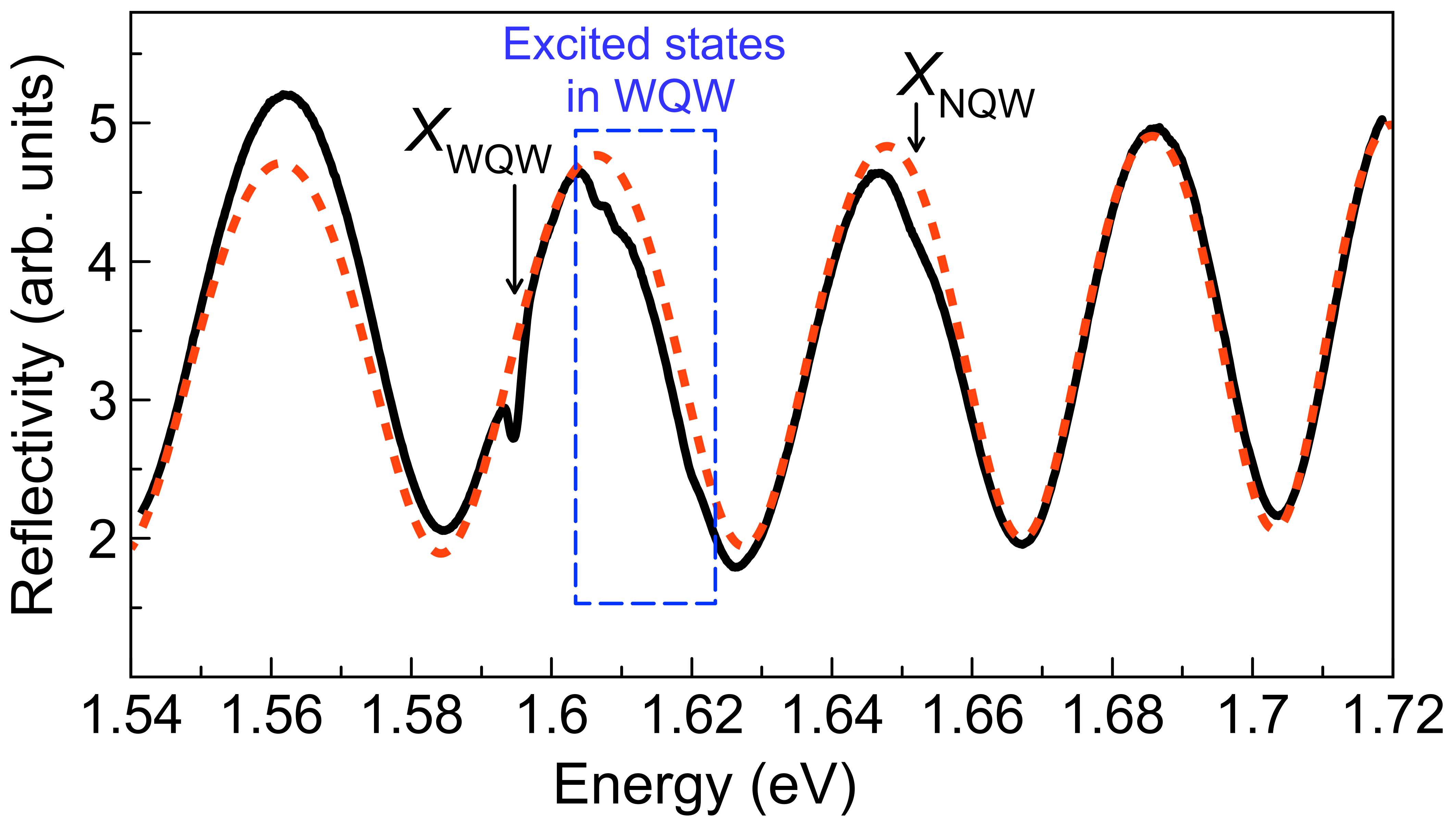}
		\caption{The reflection spectrum of the studied structure with DQW, measured in a zero magnetic field at temperature of $T=5$\,K (solid line). The vertical arrows show the excitons resonant energies $X_{\textrm{WQW}}$ and $X_{\textrm{NQW}}$ in the wide and narrow wells, the region of excited states in the wide well is highlighted by a dashed rectangle. An incandescent lamp was used as the light source. The spectrum is recorded using a 0.5-m spectrometer interfaced with a charge-coupled-device detector. The dashed curve represents the reflection spectrum calculated using the Eq.~(\ref{eq:R}).}
		\label{fig:reflection}
	\end{figure}
	
	Photoluminescence (PL) from the NQW is not observed due to the rapid tunneling of photoexcited charge carriers into the wide well. Therefore, to determine the exciton energy level in the narrow well, we used the reflection spectrum of the structure, which is shown by a solid line in Fig.\,\ref{fig:reflection}. The exciton resonance energies in the narrow and wide wells, $X_{\textrm{NQW}}$ and $X_{\textrm{WQW}}$, are shown by vertical arrows, the region of excited states in the wide well is highlighted by a dashed rectangle (for details see Ref.\,\cite{RF2024}). Strong oscillations in the reflection spectrum are caused by interference with the beam reflected from the GaAs substrate, see Appendix B for details.	
	
	\subsection{Experimental setup.}
	
	\begin{figure}[b]
		\includegraphics[width=0.85\linewidth]{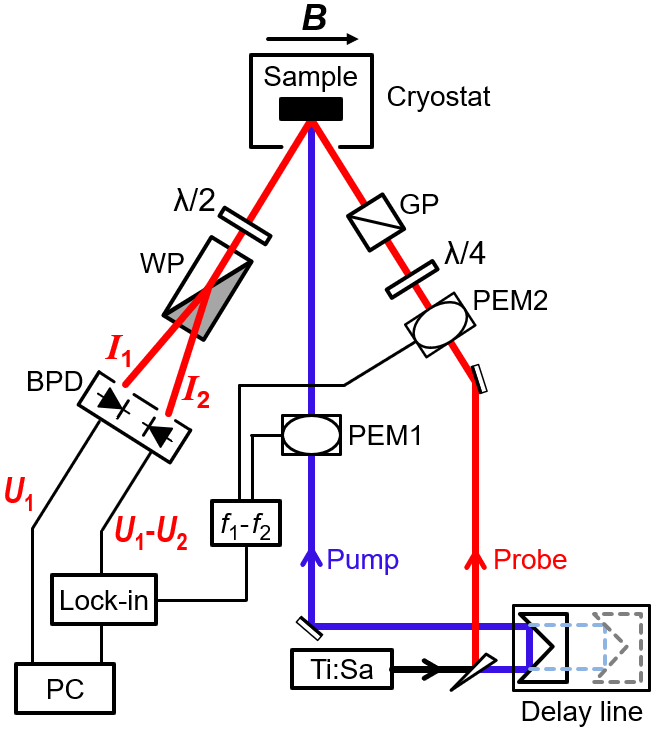}
		\caption{The scheme of experimental setup for the time-resolved spin Kerr effect measurement in pump-probe mode. Ti:Sa is the tunable Ti:Sa laser, producing 1.5\,ps pulses with the repetition rate of 80\,MHz, GP is the Glan prism, $\lambda/2$ and $\lambda/4$ are the half- and quarter-wave plates, WP is the Wollaston prism, BPD is the balanced photodetector, PEM1 and PEM2 are the photoelastic modulators, operating with frequencies $f_1=42$\,kHz and $f_2=34$\,kHz, the block ($f_1$-$f_2$) forms the reference signal for the lock-in detector with the differencial frequency ($f_1$-$f_2$), PC is the personal computer, $B$ is the external magnetic field. Electrical signals $U_1\propto I_1$, $U_2\propto I_2$.}
		\label{fig:setup}
	\end{figure}
	
	We use time-resolved degenerate pump-probe Kerr rotation (TRKR) technique~\cite{Awschalom1998, Awschalom2002} to study the optically excited spin dynamics in DQW. A schematic diagram of our experimental setup is shown in Fig.\,\ref{fig:setup}. The studied samples are placed in a closed cycle cryostat at a temperature 5\,K. The pulsed emission from a tunable mode\nobreakdash-locked Ti:Sa laser (pulse duration of 1.5\,ps with the repetition rate 80\,MHz) is split into the pump and probe beams. The pump and probe beams are directed along the normal to the sample surface and at an angle $\approx$\,$15\degree$ to it, respectively. In order to create spin polarization in the sample, the pump beam is circularly polarized. Using a tunable Ti:Sa laser allows us to perform resonant excitation of the ground exciton states in narrow or wide wells. The powers of the pump and probe beams are 20 and 5\,mW, respectively. An external magnetic field $B$ is directed perpendicular to the exciting beam (Voigt geometry).
	
	The spin Kerr effect is the rotation the linear polarization plane of the reflected probe beam by an angle $\theta$, with the magnitude controlled by spin polarization in the sample induced by the pump beam~\cite{AronIvch, Ivch2007}. For a small Kerr effect ($\theta\ll1$\,rad), which is just the case in our experiments,
	\begin{equation}
		\label{eq:theta(I)}
		\theta=\frac{(I_1-I_2)}{2(I_1+I_2)}\approx\frac{I_1-I_2}{4I_1},
	\end{equation}
	where $I_1$ and $I_2$ are the intensities of the reflected beam components linearly polarized at angles $+45\degree$ and $-45\degree$ to the incident light polarization plane. These are simultaneously recorded by photodiodes 1 and 2, comprising the balanced photodetector (BPD). Since the angle $\theta$ is small, then $I_2\approx I_1$, and the total intensity of the reflected beam is $(I_1+I_2)\approx2I_1$. To increase the measurement sensitivity, the helicity of the pump beam is modulated between $\sigma^+$ to $\sigma^-$ at the frequency of $f_1=42$\,kHz by the photoelastic modulator 1 (PEM1), while the intensity of linearly polarized probe beam is modulated by the PEM2, operating as an alternating quarter-wave plate with the frequency of $f_2=34$\,kHz together with a quarter-wave plate $\lambda/4$ and the Glan prism. The frequency of the reference signal of the lock-in detector equals $(f_1-f_2)$.
	
	\subsection{Our previous experiments of detecting Kerr effect induced by interwell exchange interaction of electrons.}

\begin{figure}[t]
		\includegraphics[width=0.9\linewidth]{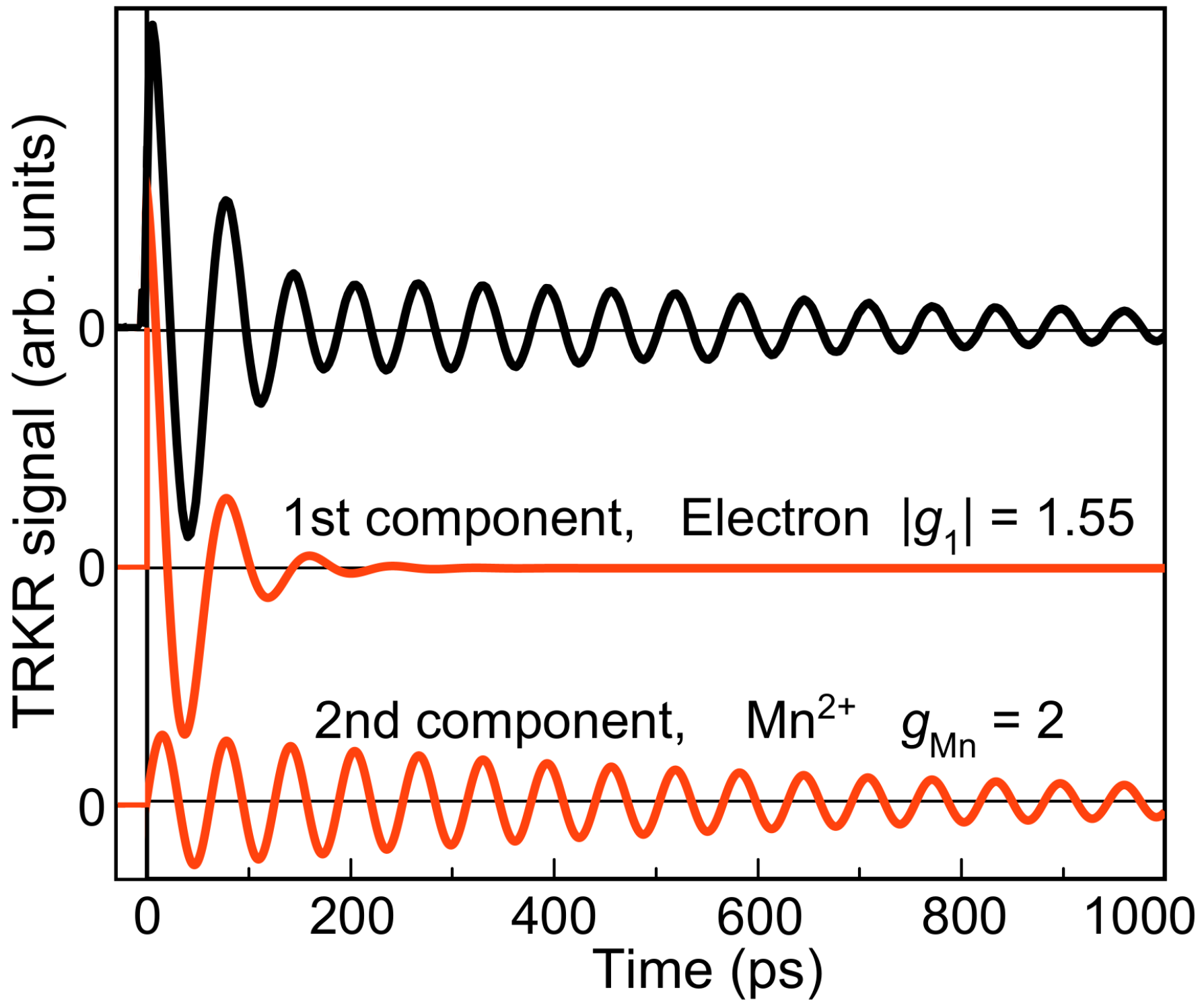}
		\caption{The Kerr rotation signal (black line) measured in the NQW in a transverse magnetic field (Voigt geometry) $B=0.54$\,T at $T=5$\,K, and its approximation by two components (red lines) shifted vertically for clarity. The pumping quantum energy  is $\hbar\omega=1.656$\,eV. }
		\label{fig:TRKR} 
	\end{figure}

	The TRKR signal measured at resonant excitation of the NQW exciton in a magnetic field in Voigt geometry is shown in Fig.\,\ref{fig:TRKR} by the black line. It is described by the sum of two exponentially decaying components (red lines), fast ($\tau_1$\,$\approx$\,47\,ps) and slow ($\tau_2$\,$\approx$\,700\,ps), of the form A$\cdot$$\exp(-t/\tau)$$\cdot$$\cos(\omega t+\varphi)$, where $\tau$ is the spin dephasing time, $\omega=|g|\mu_BB/\hbar$ is the Larmor frequency of spin precession in the magnetic field $B$, $g$ is $g$-factor, $\varphi$ is phase~\cite{Proximity, PSS2023}. The $g$-factor of the slow component equals 2, indicating that this component is caused by the Larmor precession of the Mn ions' magnetization. The presence of a fast component is unexpected, since its short coherence time is a typical feature of photoexcited excitons, whereas photoexcited carriers should be absent in the narrow well due to rapid tunneling into the wide well. At the same time, similar measurements at resonant excitation of excitons in the wide well showed the presence of a dominant ($>$\,99\% of the total signal) fast component in the Kerr signal, which is caused by the spin dynamics of electrons bound into excitons~\cite{JPCS, Proximity}. A detailed comparison of the fast components in the narrow and wide wells showed that, within the measurement error, their coherence times and  $g$\nobreakdash-factors are the same: $\tau\approx47$~ps, $|g|\approx1.53$~\cite{PSS2023}. This strongly suggests that the fast component in the narrow well is due to the spin polarization of electrons in the wide well, excited under resonant optical pumping of excitons in the narrow well.
	
	\begin{figure}[b]
		\includegraphics[width=0.85\linewidth]{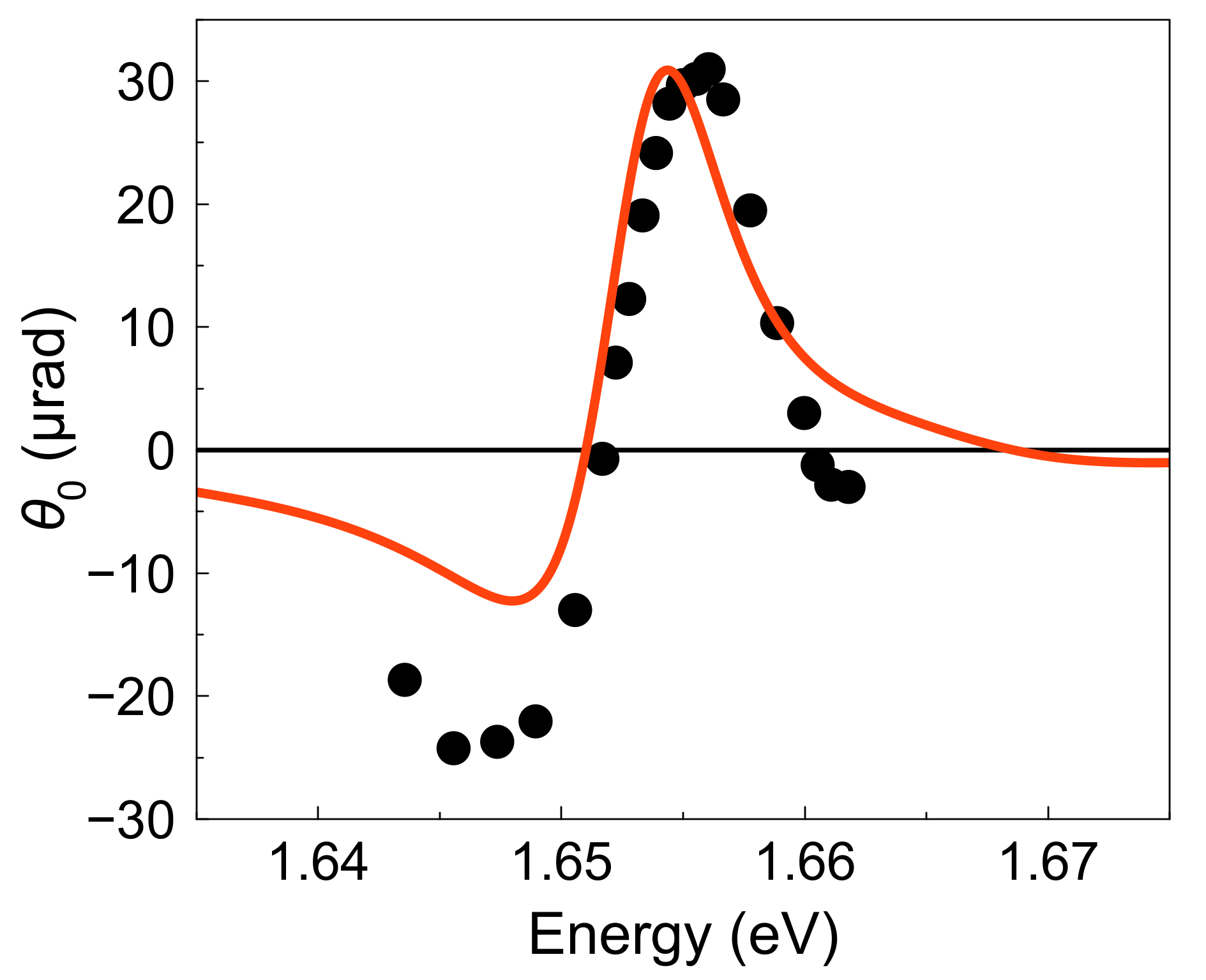}
		\caption{Spectral dependence of the amplitude of oscillations of the Kerr effect fast component occurring under resonant pumping of exciton in Cd$_{0.98}$Mn$_{0.02}$Te QW in a magnetic field $B=0.54$\,T in Voigt geometry at $T=5$\,K. Filled circles are the experimental data, solid curve is the result of calculation.} 
		\label{fig:electron_KR}
	\end{figure}
	
	The spectral dependence of the Kerr signal fast component oscillations amplitude in the narrow well (shown in Fig.\,\ref{fig:electron_KR}) has a pronounced resonant character. This excludes the possibility of a nonresonant contribution from the spin polarization of the WQW electrons. Hence, the effect is caused by the interaction of spin-polarized electrons in the wide well and electrons bound into excitons in the narrow well. More specifically, it must be the exchange component of the interaction, as the direct Coulomb interaction between excitons does not yield spin-flip terms~\cite{Gribakin2021}.
	
	\section{Theoretical estimation of electron exchange interaction}
	
	To describe the electron exchange between the narrow- and wide-well exciton ground states, one should calculate the corresponding exchange integral~\cite{landau_lifshitz_QM}. For accurate results, one needs to obtain appropriate wave functions, the quality of which can be verified by comparing the theoretical energy spectrum with the experimental one obtained from reflectivity spectra. A distinctive feature of the structure under study is that the narrow well is doped with Mn$^{2+}$ ions. This provides an additional channel for comparison with experiment, as the behavior of the energy eigenvalues as functions of the magnetic field strongly depends on the wave functions' overlap with the narrow well.
	\begin{figure}[b]
		\centering
		\includegraphics[width=1.0\linewidth]{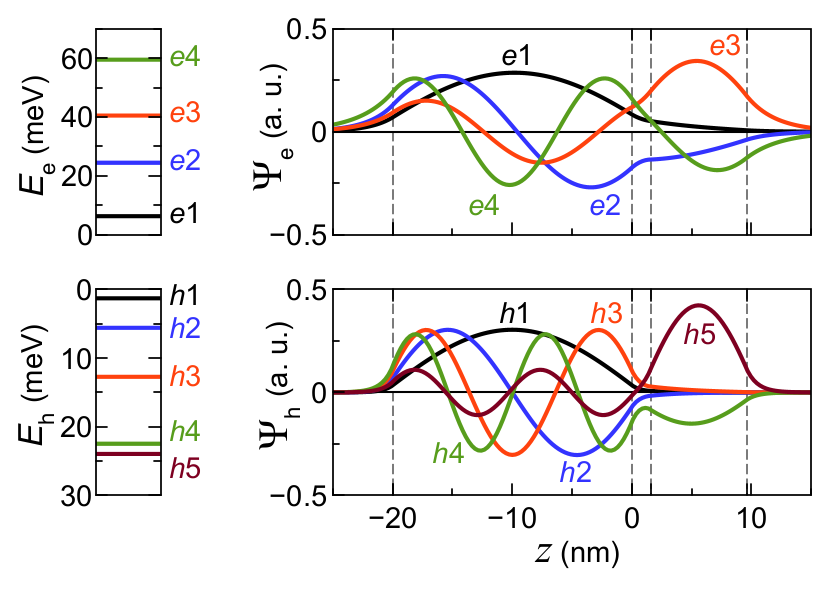}
		\caption{The energy spectrum and wave functions of the electron (top) and hole (bottom) along the growth axis of the double QW calculated taking into account the Coulomb interaction in the exciton for the spacer thickness $d=5$\,MLs. The colors of the wave functions correspond to the colors of the energy states. The borders separating the quantum wells and barriers are shown by vertical dashed lines on the right panels. The confinement energies of electrons and holes given in the panels on the left are measured from the conduction and valence band extrema of CdTe, respectively. }
		\label{fig:WFs}
	\end{figure}
	
	The problem of calculating an energy spectrum of a heterostructure is a typical one in semiconductor physics. Assuming the envelope function approximation and neglecting the difference in effective mass between the well and barrier materials, we take the electron-hole Hamiltonian
	
	\begin{equation}
		\begin{aligned}
			\label{eq:e-h_hamiltonian}
			\hat{H}_{e-h} = \frac{\hbar^2}{2 m_{e}}\nabla_{e}^{2} + \frac{\hbar^2}{2 m_{h}}\nabla_{h}^{2} + V_{e}(z_{e}) + V_{h}(z_{h})\\ + V_{e-h}\left(|{\bf r}_{e} - {\bf r}_{h}| \right),
		\end{aligned}
	\end{equation}
	where $m_{e(h)}$ is the electron (heavy hole) mass, $V_{e(h)}(z_{e(h)})$ is the quantum well potential along the growth direction, and $V_{e-h}\left(|{\bf r}_{e} - {\bf r}_{h}| \right) = e^2 / |{\bf r}_e - {\bf r}_h|$ is the Coulomb interaction of the electron and hole. The light hole is neglected due to its large confinement energy. In Eq.\,(\ref{eq:e-h_hamiltonian}) we do not write out the in-plane/out-of-plane mass anisotropy, but it is accounted for in the calculations.
	
	Converting Eq.\,(\ref{eq:e-h_hamiltonian}) to a center-of-mass coordinate system and separating the movement of the exciton as a whole, the problem is reduced to a three-dimensional one with the Hamiltonian $\hat{H}_{e-h}(\rho, z_{e}, z_{h})$, where $\rho$ is the in-plane distance between the electron and hole. This Hamiltonian is then diagonalized using the exciton basis functions
	\begin{equation}
		\psi_{n_{e} n_{h}nL}(\rho, z_{e}, z_{h}) = \phi_{e,n_{e}}(z_{e}) \phi_{h,n_{h}}(z_{h}) \Phi_{nL}(\rho),
	\end{equation}
	where $\phi_{e,n_{e}}(z_{e})$ and $\phi_{h,n_{e}}(z_{h})$ are the eigenfunctions of the electron (hole) with principal quantum number $n_{e}$ ($n_{h}$) in the quantum well potentials along the growth direction, and $\Phi_{nL}(\rho)$ are the radial wave functions of a strictly two-dimensional exciton with principal quantum number $n$ and azimuthal quantum number $L$, which have an analytical form~\cite{yang_ching1991pra_2D_hydrogen}. The radial functions are parametrized by the effective Bohr radius $a_B$. Different values were tested, but the basis $\{\psi_{n_{e} n_{h} nL}\}$ with $a_B=5$\,nm was found to be adequate, producing solutions with very little state-mixing. We have identified the narrow and wide well ground states are those based on the $e1h1$ and $e3h5$ states (Fig.\,\textcolor{red}{\ref{fig:WFs}}). To optimize agreement with experiments, we tuned the quantum well potentials slightly until reaching the same energy difference between the two states as in experiments, as well as reproducing the large effective $g$-factor of the narrow-well exciton ground state. We settled on well-established parameters from Refs.~\cite{kiselev_landwehr1998jcrystgrowth_CdTe_g_masses, long_stirner1997jap_CdTe_masses, Proximity}, except the band splitting of the narrow-well is taken $2$\,meV smaller than reported in Ref.~\cite{Proximity}. The band splitting is shared by the electron and hole as $0.55/0.45$. With these parameters, the energy difference between the two exciton ground states is reproduced within several meV, and the narrow-well exciton $g$-factor is within 15\% of the experimental one (experimental $g\sim180$, calculated $g\sim155$).  
	
	\begin{figure}[t]
		\includegraphics[width=0.9\linewidth]{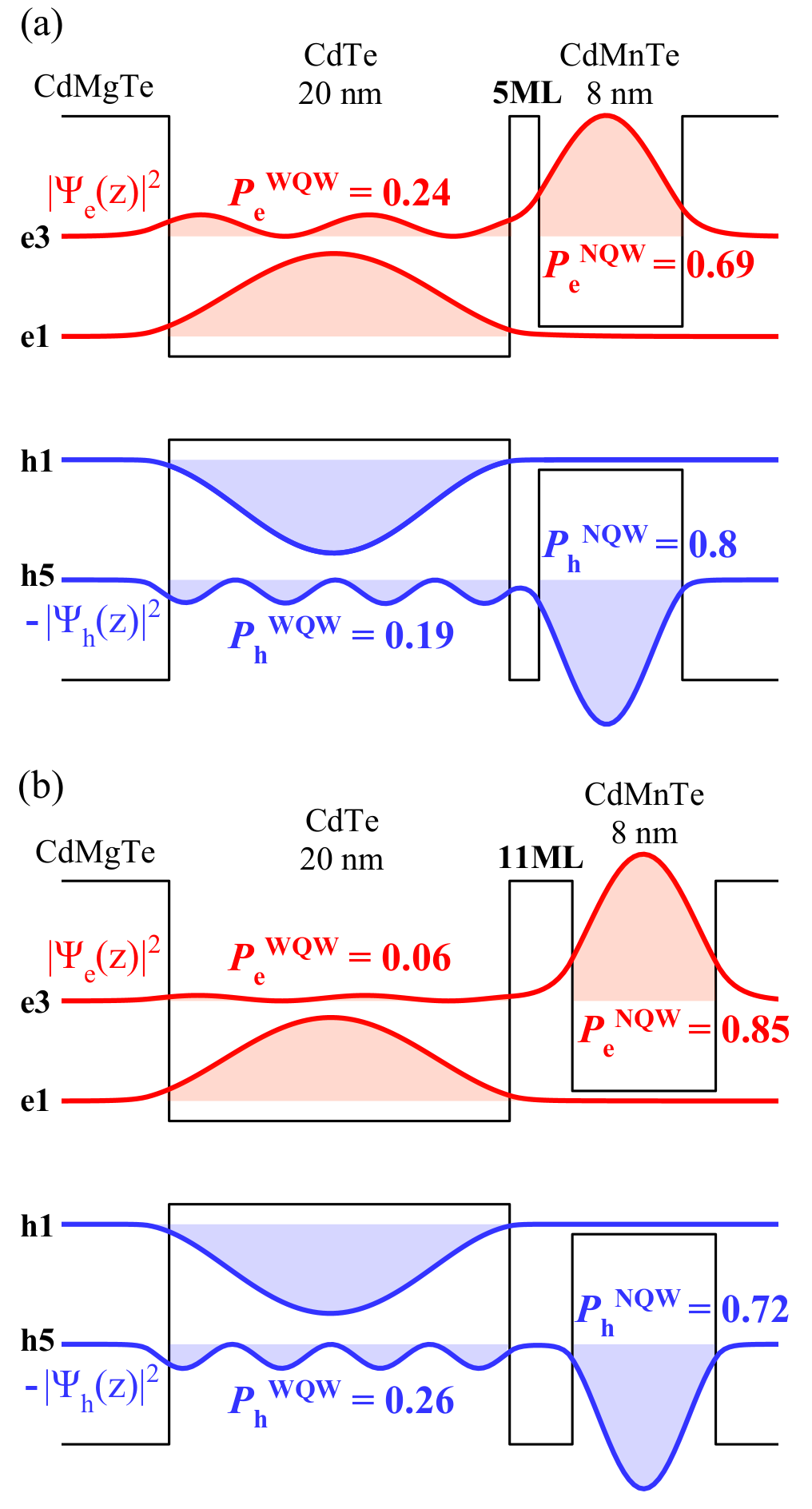}
		\caption{Squared wave functions of electrons and holes in the ground states of the wide and narrow wells, calculated for a tunnel barrier with a thickness of (a) 5~MLs and (b) 11~MLs. $P_e$ and $P_h$ are the probabilities of finding electrons and holes in wide and narrow wells in states e3 and h5.}
		\label{fig:wave functions}
	\end{figure}
	
	Having verified the quality of the exciton wave functions, we proceed to calculate the electron exchange integral between the wide- and narrow-well exciton ground states via~\cite{landau_lifshitz_QM}
	\begin{equation}
		\begin{aligned}
			\label{eq:J_exch}
			J_{\rm exch} \cdot a =\ a \int {\rm d}^{12}\  \Omega \psi_{\rm X_{\rm WQW}}(e_1, h_1) \psi_{\rm X_{\rm NQW}}(e_2, h_2)\times\\ \times V_{e_1 e_2} \psi_{\rm X_{\rm WQW}}(e_2, h_1) \psi_{\rm X_{\rm NQW}}(e_1, h_2),
		\end{aligned}
	\end{equation}
	where $a$ is exciton localization area in the NQW plane, $V_{e_1 e_2}$ is the Coulomb repulsion potential between electrons, $\psi_{\rm X_{\rm NQW (WQW)}}$ is the narrow-well (wide-well) ground state exciton wave function of the form
	\begin{equation}
		\begin{aligned}
			&\psi_{X_{\rm NQW (WQW)}}(e, h) =\\& \sum_{n_e, n_h, n, L} \mathcal{C}_{n_e n_h n L} \frac{1}{\sqrt{a}}\psi_{n_e n_h nL}(\rho, z_e, z_h)
		\end{aligned}
	\end{equation}
	with $\mathcal{C}_{n_e n_h n L}$ being the coordinates of the solutions in the $\{\psi_{n_e n_h nL}\}$ basis, and $e$ ($h$) standing for the spatial coordinates of the electron (hole).
	The normalization factor $1/\sqrt{a}$ is necessary to correctly describe the center-of-mass motion of the exciton within area $a$, which must be taken as finite but much larger than the effective radial size of the wave functions.
	Finally, the formally 12-dimensional integral has a volume element which can be written as
	\begin{gather}
		{\rm d}^{12} \Omega = {\rm d}^2 \boldsymbol{\rho}_{e_1 h_1} {\rm d}^2 \boldsymbol{\rho}_{e_2 h_2} {\rm d}^2 {\bf R}\ {\rm d}^2 \boldsymbol{\xi}\  {\rm d}z_{e_1} {\rm d}z_{h_1} {\rm d}z_{e_2} {\rm d}z_{h_2}
	\end{gather}
	with the in-plane center-of-mass coordinate ${\bf R}$ and the vector connecting the centers of mass of the excitons $\boldsymbol{\xi}$ related to the in-plane electron (hole) coordinates $\boldsymbol{\rho}_{e_{1 (2)}}$ ($\boldsymbol{\rho}_{h_{1 (2)}}$) through
	\begin{gather}
		{\bf R} = \frac{m_e(\boldsymbol{\rho}_{e_1} + \boldsymbol{\rho}_{e_2}) + m_h(\boldsymbol{\rho}_{h_1} + \boldsymbol{\rho}_{h_2})}{2M},   \\ 
		\boldsymbol{\xi} = \frac{m_e\boldsymbol{\rho}_{e_1} + m_h\boldsymbol{\rho}_{h_1}}{M} - \frac{m_e\boldsymbol{\rho}_{e_2} + m_h\boldsymbol{\rho}_{h_2}}{M}.
	\end{gather}
	
	Using Monte-Carlo integration we obtain the electron exchange integral between the narrow- and wide-well excitons:
	\begin{equation}
			\delta_e=J_{\rm exch} \cdot a=(2.1 \pm 0.3) \times 10^{-15}\ {\rm eV\,cm}^{2}.
	\end{equation}
	The uncertainty is purely numerical in nature, and is estimated as three times the variance of the integral value.
	
	The electron exchange leads to a splitting of the electron level in the narrow well $\Delta_e^e$, which is related to the constant $\delta_e$ via
	\begin{equation}
		\label{delta_e}
		\Delta_e^e=2n_e\delta_e\langle S_{\textrm{WQW}}\rangle,
	\end{equation}
	where $n_e$ and $\langle S_{\textrm{WQW}}\rangle$ are the concentration and average spin of photoexcited electrons in WQW. We found the value of $n_e\approx1.2\times10^{11}~\textrm{cm}^{-2}$ based on the average power $\langle P_{\textrm{pump}}\rangle=20$~mW, pulse repetition rate $\nu=80$~MHz and the diameter of the pump spot $d_{\textrm{pump}}\approx300~\muup$m, the light absorption coefficient $\alpha_{\textrm{CdTe}}=4\times10^4~\textrm{cm}^{-1}$ in the absorption range of CdTe~\cite{Adachi1999} and the assumption that each light quantum creates an electron-hole pair. When estimating the value of $\langle S_{\textrm{WQW}}\rangle$, we took into account that in a quantum well the circular polarization degree of luminescence with the heavy hole participation $\rho=2\langle S_{\textrm{WQW}}\rangle$~\cite{OO, SPS}. Additional measurements (not presented here) showed that under pulse pumping of an exciton in the narrow well the circular polarization degree of luminescence near maximum of exciton line in the wide well is $\rho\approx0.12$, which corresponds to $\langle S_{\textrm{WQW}}\rangle\approx0.06$ and, accordingly, $\Delta_e^e\approx30~\muup$eV.
	
	The value of the electron exchange integral defined in Eq.~(\ref{eq:J_exch}) is determined by the overlap of the respective electron wave functions. For the electron state in question, this overlap strongly depends on the spacer thickness. This is illustrated in Fig.~7(b), where the electron and hole wave functions are shown for the same nanostructure but with a thicker spacer layer of 11~MLs. A separate calculation indicates a weakening of the exchange integral by an order of magnitude as compared to the 5~MLs structure.
	
	\section{Model of spin Kerr effect induced by interwell electron exchange}
	
	Let us consider a heavy exciton in the semimagnetic (narrow) QW. A constant magnetic field applied in the plane of the QW polarizes the spins of Mn ions, which leads to the appearance of an effective exchange field, also directed in the plane of the QW, acting on the spins of charge carriers. Since a heavy hole has a $g$-factor close to zero in the QW plane, there is no noticeable splitting of the hole levels. On the other hand, electron spin sublevels corresponding to states with projections of spin +1/2 and --1/2 onto exchange field are split by energy $\Delta_{\textrm{Mn}}^e$. The Zeeman interaction of electrons with the external magnetic field can be neglected, since it is much smaller than their exchange interaction with manganese ions.
	
	The nonequilibrium spin of the electrons in the wide well, created by the pump pulse, precesses around the external field at a frequency of $\Omega_e$ and creates a rotating exchange field acting on the electrons in the narrow well, perpendicular to the external field and the exchange field of manganese. The splitting of the electron energy level in this field expressed in energy units is $\overset{\sim}\Delta\phantom{}_e^e=\Delta_e^e\exp(-t/\tau_1)$, where the multiplier $\exp(-t/\tau_1)$ describes the attenuation of the electron spin magnetization in the wide well with the characteristic time $\tau_1$. The splitting of the electron spin levels in the narrow well under the action of the total exchange field is
	\begin{equation}
		\Delta^e=\sqrt{(\overset{\sim}\Delta\phantom{}_e^e)^2+(\Delta_{\rm{Mn}}^e)^2}.
	\end{equation} 
	For simplicity, we assume the $g$-factor of electrons to be isotropic. If the exchange interaction with the wide well is weak, that is $\Delta^e_{\textrm{Mn}}\gg\Delta^e_e$, then $\Delta^e\approx\Delta^e_{\textrm{Mn}}$.
	
	The states of an electron with spin projections +1/2 and --1/2 onto the total rotating field are as follows:
	\begin{equation}
		\begin{aligned}
			\label{eq:states}
			|+1/2\rangle=&\cos{(\varphi/2)}\exp{(i\alpha/2)}|\uparrow\rangle+\\+&\sin{(\varphi/2)}\exp{(-i\alpha/2)}|\downarrow\rangle,\\\\
			|-1/2\rangle=&\cos{(\varphi/2)}\exp{(-i\alpha/2)}|\downarrow\rangle+\\+&\sin{(\varphi/2)}\exp{(i\alpha/2)}|\uparrow\rangle,
		\end{aligned}
	\end{equation}
	where $|\uparrow\rangle$ and $|\downarrow\rangle$ are the states of the electron with projections of spin +1/2 and --1/2 onto the $z$ axis of the structure, $\varphi$ is the angle of inclination of the rotating field to the $z$ axis, and $\alpha$ is the angle between the projection of the rotating field onto the plane of the structure and the external magnetic field. From simple geometric considerations, it is easy to find that
	
	\begin{equation}
		\label{eq:cos_fi}
		\cos\varphi=\frac{\overset{\sim}\Delta\phantom{}_e^e}{\sqrt{(\overset{\sim}\Delta\phantom{}_e^e)^2+(\Delta_{\textrm{Mn}}^e)^2}}\cos\Omega_et.
	\end{equation}
	\newline
	The radiative damping rate of these states due to emission of $\sigma^+$ and $\sigma^-$ photons can be written as
	\begin{equation}
		\begin{aligned}
			\label{eq:gamma}
			\Gamma_{0+}^{+1/2}=\Gamma_0|\langle\downarrow|+1/2\rangle|^2=\frac12\Gamma_0(1-\cos\varphi),\\
			\Gamma_{0+}^{-1/2}=\Gamma_0|\langle\downarrow|-1/2\rangle|^2=\frac12\Gamma_0(1+\cos\varphi),\\
			\Gamma_{0-}^{+1/2}=\Gamma_0|\langle\uparrow|+1/2\rangle|^2=\frac12\Gamma_0(1+\cos\varphi),\\
			\Gamma_{0-}^{-1/2}=\Gamma_0|\langle\uparrow|-1/2\rangle|^2=\frac12\Gamma_0(1-\cos\varphi).
		\end{aligned}
	\end{equation}
	
	The frequency- and polarization-dependent amplitude reflection coefficient of light polarized in the right ("+") and left ("\text{--}") circles near exciton resonance has the form~\cite{Ivch2007,Glaz2012}:
	\begin{equation}
		\label{eq:r_ex_+-}
		r_{\pm}^{ex}(\omega)=\frac{i\Gamma_{0,\pm}}{\omega_{0,\pm}-\omega-i(\Gamma_{0,\pm}+\Gamma_\pm)},
	\end{equation}
	where $\omega$ is the light frequency, $\omega_0$ is the exciton resonance frequency, $\Gamma_0$ and $\Gamma$ are the exciton radiative and nonradiative damping rates.
	
	Considering the nonradiative damping rate to be strong and spin-independent, $\Gamma_+=\Gamma_-=\Gamma\gg\Gamma_{0,\pm}$, which is true due to the rapid departure of charge carriers to the wide well, we find
	\begin{equation}
		\begin{aligned}
			\label{eq:r_ex}
			r_{+}^{ex}(\omega)=\frac{\dfrac i2\Gamma_0(1-\cos\varphi)}{\omega_0+\dfrac1{2\hbar}\sqrt{(\overset{\sim}\Delta\phantom{}_e^e)^2+(\Delta_{\rm{Mn}}^e)^2}-\omega-i\Gamma}+\\+\frac{\dfrac i2\Gamma_0(1+\cos\varphi)}{\omega_0-\dfrac1{2\hbar}\sqrt{(\overset{\sim}\Delta\phantom{}_e^e)^2+(\Delta_{\rm{Mn}}^e)^2}-\omega-i\Gamma},\\\\
			r_{-}^{ex}(\omega)=\frac{\dfrac i2\Gamma_0(1+\cos\varphi)}{\omega_0+\dfrac1{2\hbar}\sqrt{(\overset{\sim}\Delta\phantom{}_e^e)^2+(\Delta_{\rm{Mn}}^e)^2}-\omega-i\Gamma}+\\+\frac{\dfrac i2\Gamma_0(1-\cos\varphi)}{\omega_0-\dfrac1{2\hbar}\sqrt{(\overset{\sim}\Delta\phantom{}_e^e)^2+(\Delta_{\rm{Mn}}^e)^2}-\omega-i\Gamma},
		\end{aligned}
	\end{equation}
	and their difference
	
	\begin{equation}
		\begin{aligned}
			\label{eq:r-r}
			&r_{+}^{ex}(\omega)-r_{-}^{ex}(\omega)=\\
			&\frac{i\Gamma_0\overset{\sim}\Delta\phantom{}_e^e/\hbar\cdot\cos\Omega_et}{(\omega_0-\omega)^2-\dfrac{(\overset{\sim}\Delta\phantom{}_e^e)^2+(\Delta_{\rm{Mn}}^e)^2}{4\hbar^2}-\Gamma^2-2i\Gamma(\omega_0-\omega)}.
		\end{aligned}
	\end{equation}
	
	Reflection from the exciton produces the main contribution to the Kerr effect if the barrier bounding the QW from the side opposite the structure surface is semi-infinite. In this case, the rotation angle of linear polarization plane of the probe pulse is~\cite{Ivch2007, Glaz2012}
	
	\begin{equation}
		\label{eq:theta}
		\theta\propto Im\left(e^{i2\varphi_w}[r_+^{ex}(\omega)-r_-^{ex}(\omega)]\right),
	\end{equation}\\
	where $\varphi_w$ is the phase shift of the probe beam on the path from the middle of the well to the surface of the structure.
	
	Taking into account Eqs. (\ref{eq:cos_fi}) and (\ref{eq:r_ex}), we get
	
		\begin{widetext}
			\begin{equation}
				\label{eq:theta(w,t)1}
				\theta(\omega,t)\propto Im\left(\frac{ie^{i2\varphi_w}\Gamma_0\overset{\sim}\Delta\phantom{}_e^e/\hbar}{(\omega_0-\omega)^2-\dfrac{(\overset{\sim}\Delta\phantom{}_e^e)^2+(\Delta_{\rm{Mn}}^e)^2}{4\hbar^2}-\Gamma^2-2i\Gamma(\omega_0-\omega)}\right)\cos\Omega_et,
			\end{equation}
			where magnitude of Kerr effect is
			\begin{equation}
				\theta_0(\omega)\propto Im\left(\frac{ie^{i2\varphi_w}\Gamma_0\Delta_e^e/\hbar}{(\omega_0-\omega)^2-\dfrac{(\Delta_e^e)^2+(\Delta_{\rm{Mn}}^e)^2}{4\hbar^2}-\Gamma^2-2i\Gamma(\omega_0-\omega)}\right).
			\end{equation}
		\end{widetext}
	
	Note that in deriving (\ref{eq:theta(w,t)1}), we ignored the exchange interaction of an electron with a hole in an exciton, which, in particular, splits bright and dark excitons by the value $\Delta_0$. As shown in Appendix~A, taking into account this interaction leads to the appearance of an additional term in the difference $[r_+^{ex}(\omega)-r_-^{ex}(\omega)]$, the relative contribution of which is determined by the multiplier $\Delta_0/\hbar\Gamma$. However, due to the fact that in the narrow well under study $\Delta_0/\hbar\Gamma\ll1$, the contribution of the electron-hole exchange interaction to the equation for $\theta(\omega,t)$ can be neglected.
	
	The simple Eq.~(\ref{eq:theta}) is obtained for the case of a semi-infinite buffer layer. The reflection spectrum of our structure (Fig.\,\ref{fig:reflection}) is strongly distorted due to interference with light reflected from the substrate, which is a consequence of the finite thickness of the buffer layer. Taking into account the thickness of the barriers bounding the QW leads to a more complex expression for the Kerr angle~\cite{Gourdon2002, Gourdon2003, Ivch2007}:
	\begin{equation}
		\label{eq:theta1}
		\theta=-Im\left(\frac{r_+^{\textrm{QW}}-r_-^{\textrm{QW}}}{2r^{(0)}}\right),
	\end{equation}
	where the amplitude reflection coefficient from the QW, $r_\pm^{\textrm{QW}}$, and the polarization-insensitive amplitude reflection coefficient of the structure without taking into account the contribution of QW, $r^{(0)}$, have the form:
	\begin{equation}
		\begin{aligned}
			\label{eq:r}
			r_\pm^{\textrm{QW}}=r_\pm^{ex}\frac{4n}{(1+n)^2}e^{i2\varphi_w}A,
			~~~~~~~~~\\
			r^{(0)}=\frac{r_{0b}+r_me^{i2\varphi_{m0}}}{1+r_{0b}r_me^{i2\varphi_{m0}}},~~~~~~~~~~~
			\\
			A=\left(\frac{1+r_me^{i2(\varphi_{m0}-\varphi_{w})}}{1+r_{0b}r_me^{i2\varphi_{m0}}}\right)^2,~~
			r_{0b}=\frac{1-n}{1+n}.
		\end{aligned}
	\end{equation}
	\\
	Here, $r_\pm^{ex}$ is the exciton reflection coefficient defined in Eq.~(\ref{eq:r_ex_+-}), $n$ is the refractive index, which we assume to be the same for all layers of the structure and equal to the refractive index of the barrier from Cd$_{0.88}$Mg$_{0.12}$Te, $\varphi_{m0}=2\pi nL_{m0}/\lambda$ is the phase shift of the light wave on the $L_{m0}$ path from the interface with the substrate to the surface of the structure, $\lambda$ is the wavelength of light in vacuum, $r_m=|r_m|e^{i\varphi_m}$, $|r_m|$ and $\varphi_m$ are the modulus and phase of the reflection coefficient from the substrate.
	
	Repeating the calculations performed for the semi-infinite buffer layer, we get the following expression for the Kerr angle 
	
		\begin{widetext}
			\begin{equation}
				\label{eq:theta(w,t)2}
				\theta(\omega,t)=\frac{2n}{n^2-1}Im\left(\frac{ie^{i2\varphi_w}\Gamma_0\overset{\sim}\Delta\phantom{}_e^e/\hbar}{(\omega_0-\omega)^2-\dfrac{(\overset{\sim}\Delta\phantom{}_e^e)^2+(\Delta_{\rm{Mn}}^e)^2}{4\hbar^2}-\Gamma^2-2i\Gamma(\omega_0-\omega)}D\right)\cos\Omega_et,
		\end{equation}
		where the dimensionless complex multiplier
		
		\begin{equation}
			\label{eq:D}
			D=\frac{\left(1+|r_m|\cdot e^{i(2\varphi_{m0}+\varphi_m)}\cdot e^{-2i\varphi_w}\right)^2}{\left(1-\dfrac{n-1}{n+1}|r_m|\cdot e^{i(2\varphi_{m0}+\varphi_m)}\right)\left(1-\dfrac{n+1}{n-1}|r_m|\cdot e^{i(2\varphi_{m0}+\varphi_m)}\right)}=|D|\cdot e^{i\arg D}
		\end{equation}
		takes into account reflections from the substrate and from the crystal surface.
		
		In this case, the amplitude of the Kerr effect is
		\begin{equation}
			\label{eq:theta_0}
			\theta_0(\omega)=\frac{2n}{n^2-1} Im\left(\frac{ie^{i2\varphi_w}\Gamma_0\Delta_e^e/\hbar}{(\omega_0-\omega)^2-(\Delta_{\rm{Mn}}^e/2\hbar)^2-\Gamma^2-2i\Gamma(\omega_0-\omega)}D\right).
		\end{equation}
	\end{widetext}

When writing Eq.~(\ref{eq:theta_0}), we took into account that the amplitude of the oscillating electron field $\Delta_e^e <<\Delta_{\rm{Mn}}^e$ ($\Delta_e^e\sim10~\muup\textrm{eV}$ and $\Delta_{\rm{Mn}}^e\sim1\,$meV  as shown in Sec. III and Appendix A, respectively).

It can be seen from Eq.\,(\ref{eq:theta_0}) that using it to approximate the experimentally measured spectral dependence of the amplitude of the Kerr effect in Fig.\,\ref{fig:electron_KR} allows us to find the value $\Delta_e^e$ and estimate the value of the constant of the interwell spin-spin electron exchange $\delta_e$. However, in addition to splitting $\Delta_e^e$, Eq.~(\ref{eq:theta_0}) (taking into account Eq.~(\ref{eq:D}) for the coefficient $D$) contains 9 more parameters. Eight of them were taken from the literature data, and also found in our additional experiments described in Appendices B and C and given in Table~\ref{tab:table1}. We introduce phase shift $\varphi_w=1.15$\,rad as a constant value, which the light wave with the frequency of the exciton resonance, $X_{\textrm{NQW}}$, acquires in passing the capping layer with the thickness of 45\,nm.

\begin{table}[t]
	\begin{tabular}{ |c|c| }
		\hline
		
		\multirow{2}{*}{Parameter values} & \multirow{2}{*}{Taken from} \\
		&\\
		
		\hline
		
		\multirow{2}{*}{$n(\omega)$ in}  & \multirow{2}{*}{Ref.~\cite{n(CdMgTe)},} \\
		
		\multirow{2}{*}{Cd$_{0.88}$Mg$_{0.12}$Te barrier} & \multirow{2}{*}{Appendix B}\\
		
		&\\
		
		\hline
		
		\multirow{2}{*}{$\Delta_\textrm{Mn}^e=0.86$ meV} & \multirow{2}{*}{Appendix C} \\
		
		&\\
		
		\hline
		
		\multirow{2}{*}{$|r_m|=0.137\pm0.001$} &  \\
		
		\multirow{2}{*}{$\varphi_m=1.38\pm0.16$~rad} & \multirow{2}{*}{Appendix B} \\
		
		\multirow{2}{*}{$L_{m0}=3.573\pm0.004~\muup$m} &  \\
		
		&\\
		
		\hline
		
		\multirow{2}{*}{$\hbar\Gamma_0=91.9\pm12.1~\muup$eV} & \\
		
		\multirow{2}{*}{$\hbar\Gamma=4.6\pm0.3$~meV}  & \multirow{2}{*}{Appendix C} \\
		
		\multirow{2}{*}{$\hbar\omega_0=1.6531$~eV}  &  \\
		
		&\\
		
		\hline
		
	\end{tabular}
	\caption{The values of parameters included in Eq.\,(\ref{eq:theta_0}).}
	\label{tab:table1}
\end{table}

The result of fitting the experimental dependence $\theta_0(\omega)$ in Fig.\,\ref{fig:electron_KR} using Eq.~(\ref{eq:theta_0}) for the parameter values given in Table~\ref{tab:table1} and $\varphi_w=1.15$\,rad is shown by a solid line in Fig.\,\ref{fig:electron_KR}. The calculation accurately describes the results of the experiment, indicating that the unusual Kerr effect we discovered is indeed due to the exchange interaction of electrons in neighboring wells. The fitted value of the splitting is $(\Delta_e^e)^{exp}=13.3\pm1.3~\muup\rm{eV}$. According to Eq.~(\ref{delta_e}) and the values of the concentration and average spin of photoexcited WQW electrons estimated in Sec.~III as $n_e\approx1.2\times10^{11}~\textrm{cm}^{-2}$ and $\langle S_{\textrm{WQW}}\rangle\approx0.06$, such a splitting corresponds to the exchange interaction constant $\delta_e^{exp}\approx0.9\times10^{-15}~\textrm{eV\,cm}^{2}$. This experimentally obtained value is in agreement with the theoretical estimate $\delta_e=(2.1\pm0.3)\times10^{-15}~\rm{eV\,cm^{-2}}$ obtained in Sec.\,III, differing by a factor of about 2, which is additional confirmation of the adequacy of the Kerr effect model we proposed.

\section{CONCLUSIONS}
In a double quantum well (Cd,Mn,Mg)Te structure composed of a wide non-magnetic QW and a narrow QW doped with Mn$^{2+}$ ions, the latter being characterized by a higher-energy optical signature, the spin dynamics of the narrow QW measured under resonant pulsed excitation of the narrow QW exciton contains the wide QW electron spin dynamics, as observed via time-resolved degenerate pump-probe Kerr rotation.  By applying in-depth theoretical modeling of the Kerr effect and the energy spectrum of the studied structure, we show that this interplay arises from interwell electron exchange, which couples the photocreated exciton in the narrow QW to the electron spin ensemble residing in the lower-energy wide QW via an effective exchange field. The developed quantitative model of the magnetooptical Kerr effect explicitly incorporates this interwell exciton-electron exchange, accounting for the reflective properties of the sample as well as the radiative and non-radiative broadening of the narrow-well exciton resonance. The model allows us to reconstruct the dependence of the Kerr rotation amplitude near the narrow-well exciton resonance on the spin density of electrons in the wide QW. The accurate reproduction of the experimentally observed dependence of the Larmor-frequency beats induced by circularly polarized optical pulses showcases the accuracy of the model and the exchange interpretation. The presence of Mn$^{2+}$ ions in the narrow well allowed both an independent calibration of its optical response using the Kerr effect induced by a weak oscillating magnetic field, and an additional metric for the quantum-mechanical calculations of the structure's energy spectrum, namely the state-dependent effective $g$-factor, thus enabling a reliable extraction of the strength of the interwell exchange interaction from experiment as well as theory. The fair agreement between the theoretically and experimentally determined exchange constants ($2.1\times10^{-15}~\rm{eV\,cm^{-2}}$ and $\approx0.9\times10^{-15}~\textrm{eV\,cm}^{2}$, respectively), finally confirms that cross-barrier electron exchange governs the coupled spin dynamics in the studied structure. The obtained results establish resonant Kerr-rotation spectroscopy of narrow-well excitons as an effective tool for probing interwell electron spin interactions in both magnetic and non-magnetic semiconductor heterostructures.

\section{ACKNOWLEDGMENTS}
This investigation was supported by the Russian Science Foundation (grant No.\,23-12-00205). K.V.K. and B.F.G. acknowledge the Saint-Petersburg State University (scientific grant No.\,125022803069-4) for supporting theoretical calculations. The research of G.K. was supported by the National Science Centre (Poland) under grant No. 2012/41/B/ST3/03651.

\appendix

\section{The influence of electron-hole exchange in an exciton on the Kerr effect in the DQW under study.}

As noted in Sec.\,IV, when modeling the Kerr effect, it is necessary to take into account the exchange interaction of an electron with a hole in an exciton, which, in particular, leads to the splitting of bright and dark excitons by the value $\Delta_0$, see, for example,~\cite{Ivch2007}. In this case, the expression for heavy exciton spin levels splitting in a narrow well under the action of a total exchange field has the form:

\begin{equation}
	\begin{aligned}
		\label{eq:delta^e}
		\Delta_\pm^e&=\sqrt{(\Delta_z^e)^2+(\Delta_y^e)^2+(\Delta_x^e)^2}=\\ =&\sqrt{(\overset{\sim}\Delta\phantom{}_e^e\cos(\Omega_et)\mp\Delta_0)^2+(\Delta_{\textrm{Mn}}^e)^2+(\overset{\sim}\Delta\phantom{}_e^e\sin(\Omega_et))^2}=\\ =&\sqrt{(\overset{\sim}\Delta\phantom{}_e^e)^2\mp2\Delta_0\overset{\sim}\Delta\phantom{}_e^e\cos(\Omega_et)+(\Delta_{\textrm{Mn}}^e)^2+(\Delta_0)^2},
	\end{aligned}
\end{equation}\\
where signs “$\pm$” correspond to excitons with hole spins $\mp$3/2.\\\\

Accordingly, instead of the angle $\varphi$ (Eq.~(\ref{eq:cos_fi})), one needs to define two angles, $\varphi_+$ and $\varphi_-$, such that

\begin{equation}
	\label{eq:cos_fi2}
	\cos\varphi_\pm=\frac{\overset{\sim}\Delta\phantom{}_e^e\cos(\Omega_et)\mp\Delta_0}{\sqrt{(\overset{\sim}\Delta\phantom{}_e^e)^2\mp2\Delta_0\overset{\sim}\Delta\phantom{}_e^e\cos(\Omega_et)+(\Delta_{\textrm{Mn}}^e)^2+(\Delta_0)^2}}.
\end{equation}
\\

In this case, Eqs.~(\ref{eq:gamma}) for the radiation damping rate of exciton states take the following form:

\begin{equation}
	\begin{aligned}
		\label{eq:gamma2}
		\Gamma_{0+}^{+1/2}=\Gamma_0|\langle\downarrow|+1/2\rangle|^2=\frac12\Gamma_0(1-\cos\varphi_+),\\
		\Gamma_{0+}^{-1/2}=\Gamma_0|\langle\downarrow|-1/2\rangle|^2=\frac12\Gamma_0(1+\cos\varphi_+),\\
		\Gamma_{0-}^{+1/2}=\Gamma_0|\langle\uparrow|+1/2\rangle|^2=\frac12\Gamma_0(1+\cos\varphi_-),\\
		\Gamma_{0-}^{-1/2}=\Gamma_0|\langle\uparrow|-1/2\rangle|^2=\frac12\Gamma_0(1-\cos\varphi_-).
	\end{aligned}
\end{equation}

Accordingly, reflection coefficients and their difference equal:

	\begin{widetext}
		\begin{equation}
			\begin{aligned}
				\label{eq:r_ex2}
				r_{+}^{ex}(\omega)=\frac{\dfrac i2\Gamma_0(1-\cos\varphi_+)}{\omega_0+\omega_{e+}/2-\omega-i\Gamma}+\frac{\dfrac i2\Gamma_0(1+\cos\varphi_+)}{\omega_0-\omega_{e+}/2-\omega-i\Gamma}=i\Gamma_0\frac{\omega_0-\omega-i\Gamma+\dfrac12\omega_{e+}\cos\varphi_+}{(\omega_0-\omega)^2+\omega_{e+}^2/4-\Gamma^2-2i\Gamma(\omega_0-\omega)},
				\\
				r_{-}^{ex}(\omega)=\frac{\dfrac i2\Gamma_0(1+\cos\varphi_-)}{\omega_0+\omega_{e-}/2-\omega-i\Gamma}+\frac{\dfrac i2\Gamma_0(1-\cos\varphi_-)}{\omega_0-\omega_{e-}/2-\omega-i\Gamma}=i\Gamma_0\frac{\omega_0-\omega-i\Gamma-\dfrac12\omega_{e-}\cos\varphi_-}{(\omega_0-\omega)^2+\omega_{e-}^2/4-\Gamma^2-2i\Gamma(\omega_0-\omega)},
			\end{aligned}
		\end{equation}
		where
		$\omega_{e\pm}=\dfrac1\hbar \sqrt{(\overset{\sim}\Delta\phantom{}_e^e)^2\mp2\Delta_0\overset{\sim}\Delta\phantom{}_e^e\cos(\Omega_et)+(\Delta_{\textrm{Mn}}^e)^2+(\Delta_0)^2}$,
		
		\begin{equation}
			\begin{aligned}
				\label{eq:new_r^ex}
				&r_+^{ex}(\omega)-r_-^{ex}(\omega)=\\
				&=i\frac{\Gamma_0}\Gamma\frac{(\overset{\sim}\Delta\phantom{}_e^e/\hbar\Gamma)\left[(\omega_0-\omega)^2/\Gamma^2-\left((\overset{\sim}\Delta\phantom{}_e^e)^2+(\Delta_{Mn}^e)^2\right)/4\hbar^2\Gamma^2-1-2i(\omega_0-\omega)/\Gamma\right]\cos\Omega_et}
				{\left[(\omega_0-\omega)^2/\Gamma^2-\left((\overset{\sim}\Delta\phantom{}_e^e)^2+(\Delta_{Mn}^e)^2+(\Delta_0)^2\right)/4\hbar^2\Gamma^2-1-2i(\omega_0-\omega)/\Gamma\right]^2-(\overset{\sim}\Delta\phantom{}_e^e\Delta_0\cos{\Omega_e}t)^2/4\hbar^4\Gamma^4}-\\
				&-i\frac{\Gamma_0}\Gamma\frac{(\overset{\sim}\Delta\phantom{}_e^e\Delta_0/\hbar^2\Gamma^2)\left[(\omega_0-\omega)/\Gamma-i-\Delta_0/(4\hbar\Gamma)\right]\cos\Omega_et}
				{\left[(\omega_0-\omega)^2/\Gamma^2-\left((\overset{\sim}\Delta\phantom{}_e^e)^2+(\Delta_{Mn}^e)^2+(\Delta_0)^2\right)/4\hbar^2\Gamma^2-1-2i(\omega_0-\omega)/\Gamma\right]^2-(\overset{\sim}\Delta\phantom{}_e^e\Delta_0\cos\Omega_et)^2/4\hbar^4\Gamma^4}.
			\end{aligned}
		\end{equation}
	\end{widetext}

By the order of magnitude $\Delta_e^e\sim10~\muup\textrm{eV}$ (see theoretical estimation in Sec.\,III and experimental data in Sec.\,IV), $\hbar\Gamma\sim5~\textrm{meV}$ (see Appendix C), and $\Delta_0<260~\muup\textrm{eV}$~\cite{Kusrayev1997, Zunger}. Due to the smallness of the ratios $\Delta_e^e/\hbar\Gamma\sim10^{-3}$ and  $\Delta_0/\hbar\Gamma\sim5\times10^{-2}$, the second fraction and the term $(\Delta_e^e\Delta_0\cos\Omega_et)^2/4\hbar^4\Gamma^4$ in the denominator of the first fraction in (\ref{eq:new_r^ex}) can be ignored:
	\begin{widetext}
		\begin{equation}
			\begin{aligned}
				\label{eq:new_r^ex2}
				r_+^{ex}(\omega)-r_-^{ex}(\omega)=i\frac{\Gamma_0}\Gamma\frac{(\overset{\sim}\Delta\phantom{}_e^e/\hbar\Gamma)\left[(\omega_0-\omega)^2/\Gamma^2-\left((\overset{\sim}\Delta\phantom{}_e^e)^2+(\Delta_{Mn}^e)^2\right)/4\hbar^2\Gamma^2-1-2i(\omega_0-\omega)/\Gamma\right]}
				{\left[(\omega_0-\omega)^2/\Gamma^2-\left((\overset{\sim}\Delta\phantom{}_e^e)^2+(\Delta_{Mn}^e)^2+(\Delta_0)^2\right)/4\hbar^2\Gamma^2-1-2i(\omega_0-\omega)/\Gamma\right]^2}\cos\Omega_et.
			\end{aligned}
		\end{equation}
	\end{widetext}

According to Eq.\,(\ref{eq:Delta_X}), in the investigated semimagnetic well for the external magnetic field $B=0.54$\,T and the temperature $T=5$\,K at which the experiment was performed, the splitting $\Delta_{\textrm{Mn}}^e\approx0.86$\,meV. Therefore, $\Delta_{\textrm{Mn}}^e\gg\Delta_0$ and
\begin{equation}
	\begin{aligned}
		\label{eq:r-r_final}
		&r_{+}^{ex}(\omega)-r_{-}^{ex}(\omega)=\\
		&\frac{(i\Gamma_0\overset{\sim}\Delta\phantom{}_e^e/\hbar)\cos\Omega_et}{(\omega_0-\omega)^2-\dfrac{(\overset{\sim}\Delta\phantom{}_e^e)^2+(\Delta_{\rm{Mn}}^e)^2}{4\hbar^2}-\Gamma^2-2i\Gamma(\omega_0-\omega)}.
	\end{aligned}
\end{equation}

The Eq.~(\ref{eq:r-r_final}) coincides with the Eq.~(\ref{eq:r-r}) obtained in Sec.\,IV without taking into account the exchange interaction of an electron and a hole in the exciton. Thus, the rapid tunneling of photoexcited electrons from a semimagnetic well, leading to a large nonradiative broadening of the exciton line, as well as the dominant role of the exchange interaction with magnetic ions, make it possible to neglect the exchange interaction of an electron and a hole in a narrow well.

\section{Approximation of the experimental reflection spectrum and defining parameters $|\textbf{\textit{r}}_\textbf{\textit{m}}|,~\varphi_\textbf{\textit{m}},~\textbf{\textit{L}}_{\textbf{\textit{m}}\textbf{0}}$.}

From Fig.\,\ref{fig:reflection} it can be seen that the exciton contributions to the measured reflection spectrum of the whole structure are small and can be ignored. Then the reflectence of the structure is $R=\left|r^{(0)}\right|^2$\cite{Ivch2007} and, according to Eqs. (\ref{eq:r}),
\begin{equation}
	\begin{aligned}
		\label{eq:R}
		&R=\left(\frac{n-1}{n+1}\right)^2\times\\
		&\times\frac{1+\left(\dfrac{n+1}{n-1}\right)^2|r_m|^2-2\dfrac{n+1}{n-1}|r_m|\cos(2\varphi_{m0}+\varphi_m)}{1+\left(\dfrac{n-1}{n+1}\right)^2|r_m|^2-2\dfrac{n-1}{n+1}|r_m|\cos(2\varphi_{m0}+\varphi_m)},
	\end{aligned}
\end{equation}
where $\varphi_{m0}=2\pi nL_{m0}/\lambda$.

\begin{figure}[b]
	\centering
	\includegraphics[width=0.6\linewidth]{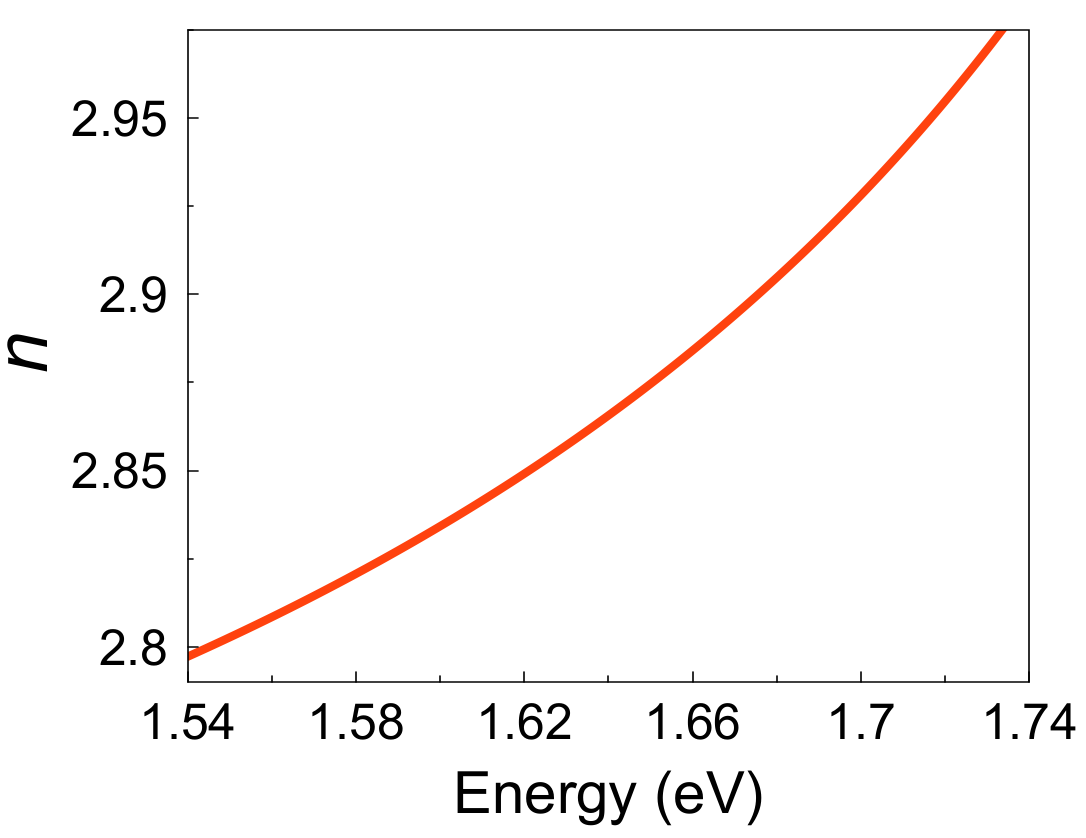}
	\caption{Refractive index spectral dependence of Cd$_{0.88}$Mg$_{0.12}$Te, based on data from~\cite{n(CdMgTe)} for helium temperature.}
	\label{fig:n}
\end{figure}

We assume that the refractive index $n$ is the same for all layers of the structure and equals to the refractive index of Cd$_{0.88}$Mg$_{0.12}$Te barriers. The spectral dependence of $n$, obtained on the base of data from the work \cite{n(CdMgTe)} and having the form $n(\hbar\omega)=\sqrt{a+b(1.24/\hbar\omega)^2/[(1.24/\hbar\omega)^2-c]}$, where $a=6.2725,~b=0.6234,~c=0.3879,\hbar\omega$ is the energy of pump quanta, is shown in Fig.\,\ref{fig:n} by a solid line.

We use Eq.~(\ref{eq:R}) to approximate the measured reflection spectrum in Fig.\,\ref{fig:reflection}. The best correspondence between the calculated (dashed curve) and experimental (solid curve) dependences is obtained at $|r_m|=0.137\pm0.001,~\varphi_m=1.38\pm0.16~\textrm{rad and}$~$L_{m0}=3.573\pm0.004~\muup\textrm{m}$. Note that the found value of $L_{m0}$ is close to the thickness of the buffer layer $\approx4~\muup\textrm{m}$ estimated during the structure growth. The discrepancy may be due to the variation in the thickness of the buffer layer in the plane of the structure, which has a wedge shape, since it was grown without substrate rotation.

\section{Measurement of the parameters ${\bf\Gamma_0,~\Gamma,~\omega_0}$ and the tunneling time of charge carriers from a narrow well using the Kerr effect in an alternating magnetic field in Faraday geometry.}

The radiative and nonradiative broadening of the exciton resonance $\Gamma_0$ and $\Gamma$ in a narrow well are important parameters determining the Kerr effect amplitude in Eq.(\ref{eq:theta_0}). It turned out to be impossible to find their values using the methods commonly used for this purpose, based on measuring the width of the exciton line in the photoluminescence or transmission spectra, in our structure with DQW. The matter is that photoluminescence from the narrow well is not observed due to the rapid tunneling of photoexcited charge carriers to the wide well, and since the studied structure is grown on a GaAs substrate, it is opaque to wavelengths in the region of exciton resonance in the narrow well. It is also difficult to use the standard method of incandescent lamp radiation reflection, since the exciton resonance in the narrow well is greatly broadened due to the short lifetime of excitons, and the reflection spectrum of the structure in Fig.\,\ref{fig:reflection} is strongly distorted as a result of interference with the light reflected from GaAs substrate.

To measure $\Gamma_0$ and $\Gamma$, we took advantage of the fact that the narrow quantum well is semimagnetic, and used the magnetooptical Kerr effect, which is effective for studying magnetization and defining exciton parameters in structures based on diluted magnetic semiconductors~\cite{Ivch2007, DMS}. The exchange interaction of charge carriers with the $d$\nobreakdash-electrons of Mn$^{2+}$ ions significantly increases the splitting of spin states of carriers and excitons in the magnetic field \cite{DMS}, which makes it possible to register magneto-optical effects even in weak magnetic fields~$\sim$~1~G.

In the magnetic field $B$ applied in Faraday geometry, the exciton resonance energy is $\hbar\omega_{0,\pm}=\hbar\omega_0\mp\Delta_X/2$, where the splitting of exciton level $\Delta_X$ consists of the splitting of electron and hole levels $\hbar\omega_e$ and $\hbar\omega_h$ induced by the magnetic field. In a semimagnetic semiconductor, the value of $\Delta_X$ is determined by the $s/p$-$d$ exchange interaction of electrons and holes with manganese ions and, in the first order in the magnetic field, equals~\cite{DMS, Scalbert}
\begin{equation}
	\begin{aligned}
		\label{eq:Delta_X}
		\Delta_X=\Delta_{\textrm{Mn}}^e+\Delta_{\textrm{Mn}}^h=~~~~~~~~~~~~~~~~~\\
		(P_e\alpha N_0-P_h\beta N_0)x_{\textrm{eff}}\frac{S(S+1)}3\frac{g_\textrm{Mn}\mu_B B}{k_B(T+T_0)},
	\end{aligned}
\end{equation}
where $\alpha N_0=0.22~\rm{eV}$ and $\beta N_0=-0.88~\rm{eV}$ are constants of the exchange interaction of electrons and holes with manganese ions, $S=5/2$ and $g_{\rm{Mn}}=2$ are spin and \textit{g}\nobreakdash-factor of the Mn ion, $T_0=35.37x/(1+2.752x)$ and $x_{\textrm{eff}}=[0.265\exp(-43.34x)+0.735\exp(-6.19x)]x$ are phenomenological parameters that take into account the antiferromagnetic exchange interaction of manganese ions. The values $T_0=0.67$ and $x_{\textrm{eff}}=0.015$ correspond to the manganese content $x=0.02$ in the semimagnetic well under study. The values of the squared overlap integrals of the electron and hole wave functions with a semimagnetic well $P_e=0.69$ and $P_h=0.8$ were obtained by numerical calculation in Sec.\,III.

In order to increase the measurement sensitivity we used an alternating magnetic field $B(t)=B_1\cos(\Omega_1t)$ in Faraday geometry and lock-in detection.\,\,In this case, using the Eq.~(\ref{eq:theta1}), we find that $\theta(\omega,t)=\theta_0(\omega)\cos(\Omega_1t)$, where, when the condition $\Gamma_0\ll\Gamma$ is fulfilled,
\begin{equation}
	\begin{aligned}
		\label{eq:theta_01}
		\theta_0(\omega)=\frac{2n}{n^2-1}Im\left(\frac{i(\Gamma_0/\hbar)e^{i2\varphi_w}\Delta_{X}(B_1)}{(\omega_0-\omega)^2-\Gamma^2-2i\Gamma(\omega_0-\omega)}D\right)
	\end{aligned}
\end{equation}
and multiplier $D$ is defined by Eq.~(\ref{eq:D}).

\begin{figure}[t]
	\centering
	\includegraphics[width=0.9\linewidth]{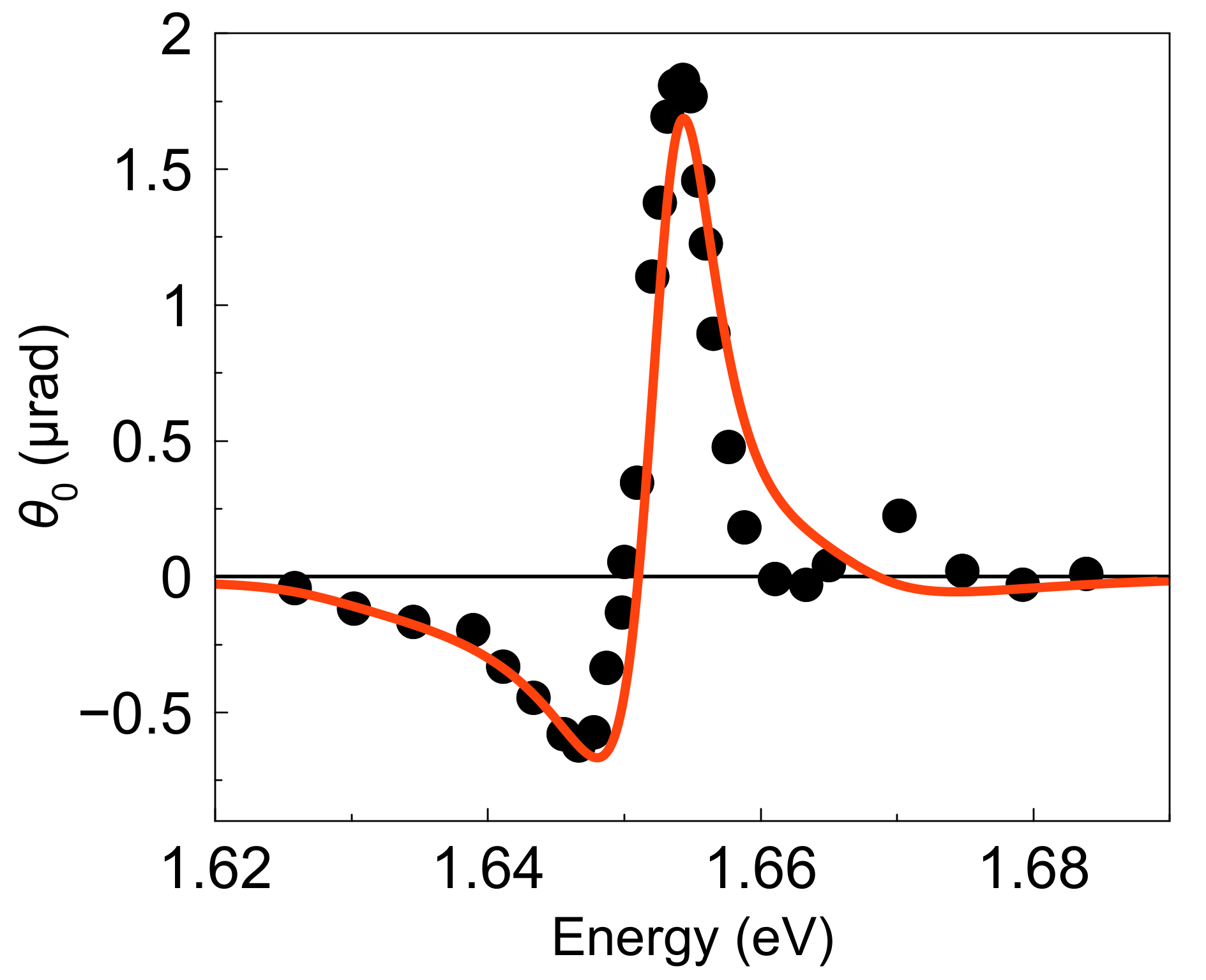}
	\caption{Experimental (circles) and calculated (solid curve) spectral dependences of the Kerr angle amplitude in the exciton resonance region in the semimagnetic QW in a longitudinal alternating magnetic field $B(t)=B_1\cos(\Omega_1t)$ at $B_1=0.8$\,G, $\Omega_1=130$~Hz and temperature $T=5$~K. Under such conditions, exciton level splitting is $\Delta_{X}(B_1)=0.72~\muup$eV.}
	\label{fig:KR_long}
\end{figure}

Spectral dependence of $\theta_0(\omega)$, measured in exciton resonance region in a semimagnetic QW at $B_1=0.8$\,G, $\Omega_1=130$\,Hz and temperature $T=5$\,K and shown as filled circles in Fig.\,\ref{fig:KR_long}, has a strongly expressed resonant character. Its approximation using the Eq.~(\ref{eq:theta_01}) (solid curve) allowed us to find that $\hbar\Gamma_0=91.9\pm12.9~\muup$eV, $\hbar\Gamma=4.6\pm0.3$~meV, $\hbar\omega_0=1.6531~$eV at $\varphi_w=1.15$~rad. With this $\Gamma_0$ to $\Gamma$ ratio, the observed resonance width is determined by nonradiative broadening, while radiative broadening sets its amplitude.

The visible width of the exciton resonance may increase due to inhomogeneous broadening caused, for example, by fluctuations of the quantum well width. However, in single semimagnetic quantum wells having similar composition and the same crystal temperature, the inhomogeneous broadening is $\hbar\Gamma_{\textrm{inh}}\sim0.7$~meV~\cite{Gamma_inh}, which is almost an order of magnitude less than the measured value of $\hbar\Gamma\approx4.6$~meV. Therefore, the influence of inhomogeneous broadening on the resonance width and the values of the parameters determined in our experiments can be ignored. It can be concluded that in the DQW structure under study a significant exciton resonance width in the narrow well is due to the rapid tunneling of charge carriers into the wide well with characteristic time $\tau_{\textrm{tun}} =(2\Gamma)^{-1}\approx0.1$\,ps.

Our results show that the magnetooptical Kerr effect makes it possible to study exciton states in structures with short nonradiative times, where traditional methods of photoluminescence, transmission, and reflection spectroscopy may be inapplicable or ineffective.

\begin{figure}[b]
	\centering
	\includegraphics[width=0.9\linewidth]{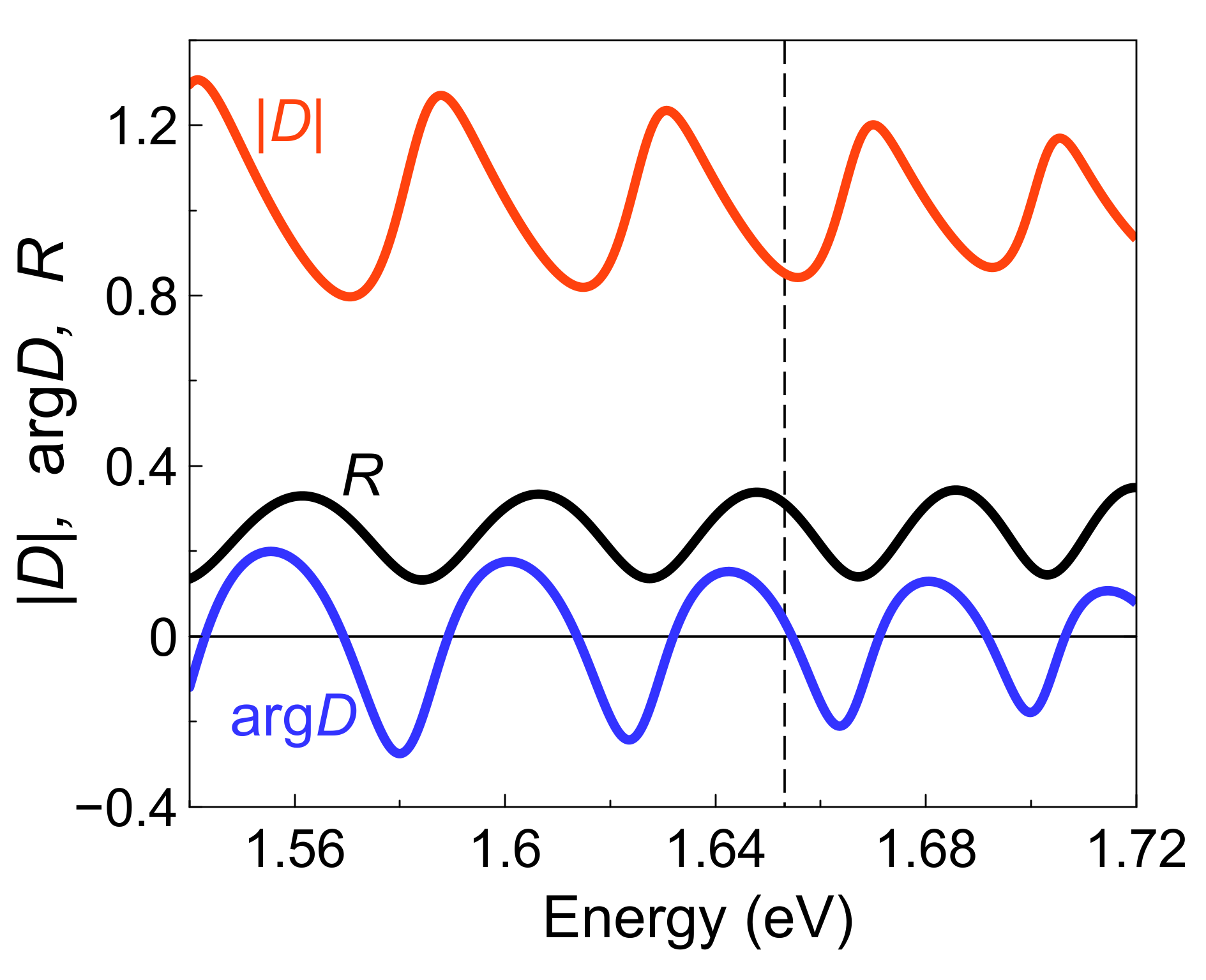}
	\caption{Spectral dependencies of $|D|,~\arg D$ and $R$ calculated using Eqs. (\ref{eq:D}) and (\ref{eq:R}) with  $|r_m|=0.137$, $\varphi_m=1.38$\,rad, $L_{m0}=3.57\,\muup$m and $\varphi_w=1.15\,$rad. Vertical dashed line shows the exciton energy $X_\textrm{NQW}$.}   
	\label{fig:D,argD,R}
\end{figure}

Using the parameter values found in Appendices~B and C, the spectral dependences of $|D|$ and $\arg D$ can be calculated and the influence of reflections from the substrate and the structure surface can be estimated. Spectral dependences of $|D|$ and $\arg D$ calculated with parameters $|r_m|=0.137$, $\varphi_m=1.38$~rad, $L_{m0}=3.57~\muup$m and $\varphi_w=1.15$~rad, are shown in Fig.\,\ref{fig:D,argD,R} together with the spectral dependence for $R$. It can be seen that the dependencies of $|D|$ and $\arg D$ change significantly, oscillating with the $R$ variation period, and $\arg D$ even changes sign. Note that in the absence of reflection from the substrate, there is no dispersion of $D$: $|D|=1$ and $\arg D=0$.

\end{document}